\documentclass[%
 reprint,
 superscriptaddress,
 amsmath,amssymb,
 aps,
 pra,
]{revtex4-2}

\usepackage{graphicx}
\usepackage{dcolumn}
\usepackage{bm}
\usepackage{physics}
\usepackage{verbatim}
\usepackage[small,tight,FIGTOPCAP]{subfigure}
\usepackage{amsmath}
\usepackage{amsfonts}
\usepackage{amssymb}
\usepackage[]{mathrsfs}
\usepackage{xcolor}
\usepackage{bbm}
\usepackage [english]{babel}

\begin{document}

\preprint{APS/123-QED}


\title{Generating high-order exceptional points in coupled electronic oscillators using complex synthetic gauge fields}

\author{José D. Huerta-Morales}
\email{jose.huerta@correo.nucleares.unam.mx}
\affiliation{Instituto de Ciencias Nucleares, Universidad Nacional Aut\'onoma de M\'exico, Apartado Postal 70-543, 04510 Cd. Mx., M\'exico}

\author{Mario A. Quiroz-Ju\'{a}rez}
\affiliation{Centro de F\'{i}sica Aplicada y Tecnolog\'{i}a Avanzada, Universidad Nacional Aut\'onoma de M\'exico, Boulevard Juriquilla 3001, 76230 Quer\'{e}taro, M\'exico}


\author{Yogesh N. Joglekar}
\email{yojoglek@iupui.edu}
\affiliation{Department of Physics, Indiana University - Purdue University Indianapolis (IUPUI), Indianapolis, Indiana 46202 USA}

\author{Roberto de J. Le\'on-Montiel}
\email{roberto.leon@nucleares.unam.mx}
\affiliation{Instituto de Ciencias Nucleares, Universidad Nacional Aut\'onoma de M\'exico, Apartado Postal 70-543, 04510 Cd. Mx., M\'exico}

\date{\today}

\begin{abstract}
Exceptional points (EPs) are degeneracies of non-Hermitian systems, where both eigenvalues and eigenvectors coalesce. Classical and quantum systems exhibiting high-order EPs have recently been identified as fundamental building blocks for the development of novel, ultra-sensitive opto-electronic devices. However, arguably one of their major drawbacks is that they rely on non-linear amplification processes that could limit their potential applications, particularly in the quantum realm. In this work, we show that high-order EPs can be designed by means of linear, time-modulated, chain of inductively coupled RLC (where R stands for resistance, L for inductance, and C for capacitance) electronic circuits. With a general theory, we show that $N$ coupled circuits with $2N$ dynamical variables and time-dependent parameters can be mapped onto an $N$-site, time-dependent, non-Hermitian Hamiltonian, and obtain constraints for $\mathcal{PT}$-symmetry in such models. With numerical calculations, we obtain the Floquet exceptional contours of order $N$ by studying the energy dynamics in the circuit. Our results pave the way toward realizing robust, arbitrary-order EPs by means of synthetic gauge fields, with important implications for sensing, energy transfer, and topology.
\end{abstract}

\maketitle


\section{\label{sec:level1}Introduction}
The past two decades have witnessed a Cambrian explosion of several experimental and theoretical investigations on non-Hermitian Hamiltonian systems that satisfy the so-called parity-time ($\mathcal{PT}$) symmetry condition \cite{Bender1998p5243,Bender1999p2201,Bender2005p277,Levait2000,HuertaMorales2016p83}. Owing to the antilinear $\mathcal{PT}$-symmetry, the spectrum of such Hamiltonians changes from real to complex-conjugate pairs as the degree of anti-Hermiticity is increased. The coherent, non-unitary evolution generated by the non-Hermitian Hamiltonian means that the norm of a state oscillates in the $\mathcal{PT}$-symmetric region, where the spectrum is real, and grows exponentially in the $\mathcal{PT}$-broken region, where amplifying (and decaying) eigenmodes are present. The study and analysis of $\mathcal{PT}$-symmetric systems across the parameter domain have triggered important theoretical predictions and experimental demonstrations in disparate areas of physics, optics, and photonics \cite{Siviloglou2009p093902,Ruter2010p192,Feng2017p752,El-Ganainy2018p11}. Specifically, the non-trivial phenomena across the $\mathcal{PT}$-transition have captured a great deal of attention. This is, in part, because the exceptional-point degeneracy--in contrast with the traditional Hermitian degeneracy---is a potentially good candidate for sensing small disturbances due to a perturbing potential~\cite{Hichem2016p042116,Ozdemir2019p783,Sakhdari2019p193901,Hodaei2017p187,Chen2017p192,OptExpress.27.37494,ROSA2021104325}. 

For Hermitian Hamiltonians, however, even when two eigenvalues become degenerate, two orthonormal eigenvectors remain. Exceptional points (EPs) are thus non-Hermitian singularities where two or more eigenvectors also coalesce
\cite{Rotter2009p153001,Muller2008p244018,Teimourpour2018}. The number of eigenvectors $n$ that collapse at the non-Hermitian degeneracy defines the order of the EP, and we call it, EP$n$. The most common case is an EP2, where a pair of eigenvalues become degenerate, and EPs of order greater than two are traditionally referred to as high-order EPs \cite{Heiss2012p444016,Quiroz-Juarez2019p862,Zhang2020p033820,Nada2020}. The literature on the design and realization of high-order EPs has greatly contributed to the development of this research field, most of them aimed at enhancing the response of open physical systems \cite{PhysRevLett.112.203901,PhysRevLett.117.110802} because the dimensionless mode-splitting $\Delta\omega$ in response to a dimensionless perturbation $\delta\ll 1$ at an EP$n$ is given by $\Delta\omega(\delta)\propto\delta^{1/n}\gg\delta$. Put simply, the sensitivity to perturbations increases as the order of the EP increases. For instance, a 1\% perturbation results in a 1\% response in a Hermitian system, a 10\% response at an EP2, and a 30\% response at an EP3.

Recent theoretical and experimental studies have focused on diverse $\mathcal{PT}$-symmetric platforms to realize EPs of arbitrary order. Examples include waveguide-arrays \cite{El-Ganainy2018p11,Ruter2010p192,Moiseyev2021p033518}, micro-cavities \cite{Peng2014p394,Hodaei2017p187,Chen2017p192}, opto-mechanical systems \cite{Jing2017p3386,JaramilloAvila2020,Xiong2021p063508}, quantum optical circuits \cite{Quijandria2018p053846,Zhihao2020p022039}, coupled acoustic resonators \cite{Shi2016p11110,Ding2016p021007,Wang2019p832}, and electronic circuits \cite{Sakhdari2019p193901,Schindler2011p040101,Lin2012p050101,Kazemi2019,Kazemi2022}. The latter, in recent years, have emerged as a powerful platform for simulating topological and non-Hermitian phenomena~\cite{PhysRevLett.126.215302,PhysRevLett.128.065701,Cao2022}. In particular, we have shown that both gain and loss can be implemented in a single LC oscillator by means of synthetic gauge fields, thereby creating static and Floquet EP2 landscape \cite{Quiroz-Juarez2021parXiv}. Extending this approach to higher-dimensional EPs is, however, non-trivial. 

\begin{figure*}
\includegraphics[width=1\linewidth]{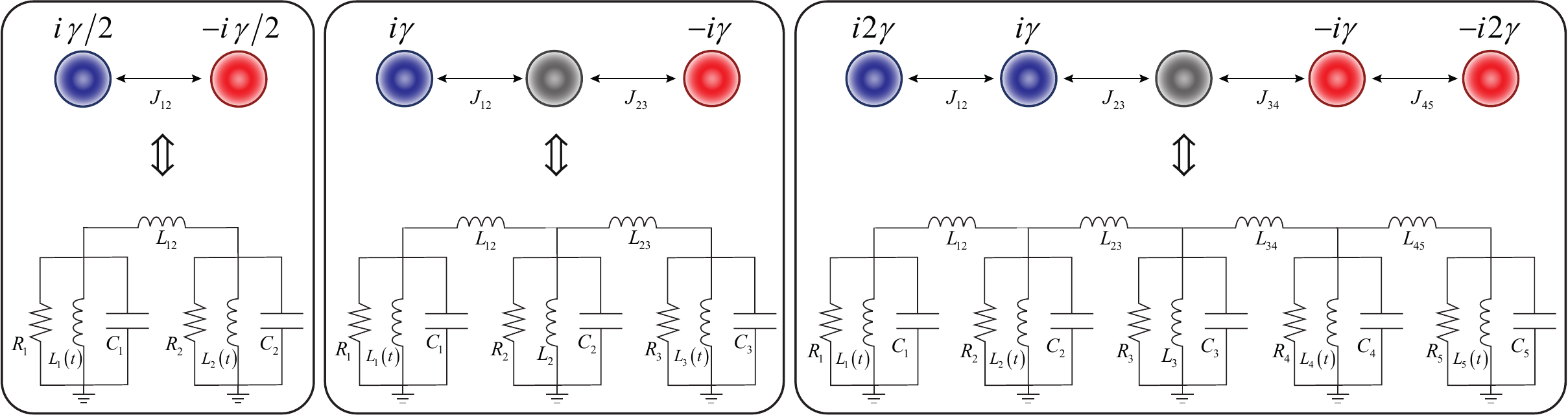}
\caption{\label{fig:sitios-circuitos} \textbf{Schematic representation of the equivalence between sites and electronic circuits in a tight-binding lattice model}. The top row shows $\mathcal{PT}$-symmetric perfect-state-transfer lattice with $N=2,3,5$ sites respectively. Its Hamiltonian $H_N(\gamma)=\kappa J_y+i\gamma J_z$ is $N$ dimensional. Bottom row shows $N$ inductively coupled RLC circuits with $2N$ dynamical variables,  exhibiting synthetic gain and loss by means of time-dependent inductance within each RLC box.}
\end{figure*}

In this work, we show through analytical and numerical methods that it is possible to engineer high-order EPs in an inductively-coupled RLC-circuit tight-binding lattice. We use the $J_y$ array--a tight binding lattice with non-uniform couplings that has equidistant eigenvalues~\cite{Joglekar2013,PhysRevA.87.022303,NatCommun.7.11027} -- along with a gain-loss profile that mimics the $J_z$ array to realize higher-order EPs~\cite{Teimourpour2018}. Specifically, we implement the non-trivial features of $\mathcal{PT}$-symmetry with synthetic gain and loss through the temporal variation of the inductances in each oscillator~\cite{Kazemi2019,Kazemi2022,Quiroz-Juarez2021parXiv}. Our results suggest that dynamically tunable synthetic electronics, with $\mathcal{PT}$-symmetry implemented through a complex gauge field, can be used to simulate higher-order EPs. 

The paper is organized as follows. In Sec.~\ref{sec:model} we present the formalism that maps the Kirchoff-law equations for currents and voltages in a chain of $N$ inductively coupled RLC oscillators into a Schrodinger-like equation with a $N_e\equiv(3N-1)$ dimensional Hamiltonian, and show how a time-dependent, non-unitary change of basis can lead to gain and loss. In Sec.~\ref{sec:floquet} we demonstrate that the time-modulation of specific components of those oscillators can create EP$N$ contours. Section~\ref{sec:disc} provides a brief discussion and conclusions. The explicit forms of the Hamiltonians for $N=3,4,5$ are given in the Appendix.


\section{The Model}
\label{sec:model}

Let us consider a set of $N$ RLC circuits connected by $(N-1)$ coupling inductors (Fig.~\ref{fig:sitios-circuitos}). Their dynamics are governed by the following first-order equations~\cite{PhysRevResearch.3.013010} 
\begin{subequations}
\begin{eqnarray} 
	\frac{d V_{n}}{dt}&=&\frac{1}{C_{n}}\left[-\frac{V_{n}}{R_{n}}-I_{n} - I_{x,n} + I_{x,n-1}    \right], \label{eq1_1} \\
	\frac{d I_{n}}{dt}&=&\frac{1}{L_{n}}V_{n},  \label{eq1_2} \\
	\frac{dI^x_{n}}{dt}&=&\frac{1}{L_{x,n}}\left(V_{n} - V_{m}\right). \label{eq_3}
\end{eqnarray}
\end{subequations}
These equations arise from the Kirchhoff laws. Here, $V_{n}(t)$ is the voltage in the capacitor $C_{n}$, $I_{n}(t)$ is the current across the inductor $L_{n}$, $R_{n}$ is the resistance in the {\it n}th oscillator, $L_{x,n}$ denotes the inductor coupling the {\it n}th RLC box to the 
({\it n}+1)th box, and $I_{x,n}(t)$ is the current flowing across it. We can write the $N_e\equiv(3N-1)$ Eqs.~(\ref{eq1_1})-(\ref{eq_3}) in a compact form as
\begin{eqnarray}  \label{Eq.Shrodinger0}
	i \frac{d}{dt} \ket{\phi \left(t\right)}=H \left(t\right) \ket{ \phi \left(t\right)},
\end{eqnarray}
where the $N_e$-dimensional ``state-vector'' is 
\begin{eqnarray}
	\ket {\phi \left(t\right)}=\begin{pmatrix}
		V_1,\cdots,V_N,I_1,\cdots,I_N,I_{x,1},\cdots,I_{x,N-1}
	\end{pmatrix}^{T}
\end{eqnarray} 
$H\left(t\right)$ is a non-Hermitian, $N_e\times N_e$ matrix with purely imaginary entries. To specify its general structure, we define ancillary matrices $\mathbb{C}=\textrm{diag}(C_1,\cdots,C_N)$, $\mathbb{L}=\textrm{diag}(L_1,\cdots,L_N)$, $\mathbb{G}_{RC}=\textrm{diag}(1/R_1C_1,\cdots,1/R_NC_N)$, and $\mathbb{L}_x=\textrm{diag}(L_{x,1},\cdots,L_{x,N-1})$. Additionally, we also define an $N\times(N-1)$,  almost skew-symmetric matrix $\mathbb{S}$ with entries $\mathbb{S}_{ab}=\delta_{ab}-\delta_{a,b+1}$. In terms of these matrices, $H$ can be written as
\begin{equation}
\label{eq:h}
H=i\left[\begin{array}{ccc}
-\mathbb{G}_{RC} & -\mathbb{C}^{-1} & -\mathbb{C}^{-1}\mathbb{S}\\
\mathbb{L}^{-1} & 0 & 0\\
\mathbb{L}_x^{-1}\mathbb{S}^\dagger & 0 & 0
\end{array}\right].
\end{equation}

Since the state vector $|\phi(t)\rangle$ has entries with different engineering dimensions, so does the matrix $H$. To clarify its underlying symmetry properties, it is useful to consider the ``square-root-of-energy'' basis. The energy in the $N$-node circuit is given by 
\begin{align}
    E(t)&=\frac{1}{2}\sum_{n=1}^N\left[C_n V_n^2+L_nI_n^2\right]+\frac{1}{2}\sum_{m=1}^{(N-1)}L_{x,m}I^2_{x,m},\nonumber \\
    &=\langle\phi(t)|A|\phi(t)\rangle,\label{eq:energy}
\end{align}
where the positive, $N_e$ dimensional bilinear-form matrix is given by $A=\textrm{diag}(\mathbb{C},\mathbb{L},\mathbb{L}_x)/2$. We define $|\psi(t)\rangle=A^{1/2}(t)|\phi(t)\rangle$ so that norm of $|\psi(t)\rangle$ encodes the circuit energy, $\langle\psi(t)|\psi(t)\rangle=E(t)$. Note that all entries in the state vector $|\psi(t)\rangle$ have units of $\sqrt{\textrm{Joules}}$. It is straightforward to show that $\ket{ \psi \left(t\right)}$ satisfies a Schrödinger-like equation
\begin{equation}
	i \frac{d}{dt} \ket{ \psi \left(t\right) }=(H_0+\Gamma)\ket{\psi \left(t\right)}=H_\textrm{cir}(t)|\psi(t)\rangle,
\end{equation}
where the effective circuit Hamiltonian $H_\textrm{cic}=H_0+\Gamma$ has two components. The first component $H_0$ is given by 
\begin{eqnarray}
    H_{0}&=& \sqrt{A}H\frac{1}{\sqrt{A}}=i\left[\begin{array}{ccc}
    -\mathbb{G}_{RC} & -\mathbb{W} & - \frac{1}{\sqrt{\mathbb{C}}}\mathbb{S}\frac{1}{\sqrt{\mathbb{L}_x}}\\
    \mathbb{W} & 0 & 0\\
    \frac{1}{\sqrt{\mathbb{L}_x}}\mathbb{S}^\dagger\frac{1}{\sqrt{\mathbb{C}}} & 0 & 0
    \end{array}\right] \label{Eq_H0G} 
\end{eqnarray}
where $\mathbb{W}=\textrm{diag}(\omega_1,\cdots,\omega_N)$ is a diagonal matrix with frequencies of individual oscillators $\omega_k=1/\sqrt{C_kL_k}$. When there is no dissipation in each $RLC$ circuit, i.e. $\mathbb{G}_{RC}=0$, the matrix $H_0$ becomes Hermitian and the corresponding unitary evolution of the state $|\psi(t)\rangle$ signals the conservation of total energy in the circuit. When $\mathbb{G}_{RC}>0$, this anti-Hermitian piece of $H_0$ encodes the Joule dissipation. The second component of $H_\textrm{cir}$ is given by 
\begin{equation}
\Gamma \left(t\right) = i\frac{1}{\sqrt{A}}\frac{d}{dt}\sqrt{A}= i \frac{d}{dt}\ln\sqrt{A\left(t\right)}. 
    \label{Eq_GammaG} 
\end{equation} 
If the change of basis matrix $\sqrt{A(t)}$ is unitary, as is 
typically the case, its spatiotemporal variations give rise to a Hermitian $\Gamma(t)$ since the logarithm of a unitary matrix is an anti-Hermitian matrix. However, as our change of basis matrix $A=\textrm{diag}(\mathbb{C},\mathbb{L},\mathbb{L}_x)/2$ is not unitary and always real, it leads to an anti-Hermitian, gain-loss term $\Gamma(t)$, as defined in Eq.~(\ref{Eq_GammaG}). 

The parity operator exchanges the node $n$ with its mirror symmetric node $\bar{a}=N+1-a$. Therefore, it is given by $\mathcal{P}=\textrm{diag}(\Pi_N,\Pi_N,-\Pi_{N-1})$ where $\Pi_k=\Pi_k^{-1}=\Pi_k^\dagger$ is the antidiagonal matrix of size $k$ with unit entries. The time-reversal operator, in addition to the complex-conjugation operation $*$, reverses the sign of each current: $\mathcal{T}=\textrm{diag}(\mathbbm{1}_N,-\mathbbm{1}_N,-\mathbbm{1}_{N-1})*$. Thus, the antilinear $\mathcal{PT}$ operator is given by the following $N_e$-dimensional, block-diagonal matrix
\begin{equation}
\label{eq:ptne}
\mathcal{PT}=\textrm{diag}(\Pi_N,-\Pi_N,\Pi_{N-1})*=\mathcal{U}*,
\end{equation}
where $\mathcal{U}$ denotes the $N_e$-dimensional real, unitary matrix.  
By imposing the constraint that $H_0$ is $\mathcal{PT}$-symmetric, we get 
\begin{align}
    &\Pi_N\mathbb{G}_{RC}\Pi_N=-\mathbb{G}_{RC}=0,\\
    & \Pi_N\mathbb{W}\Pi_N=\mathbb{W},\\
    & \Pi_N\frac{1}{\sqrt{\mathbb{C}}}\mathbb{S}\frac{1}{\sqrt{\mathbb{L}_x}}=\frac{1}{\sqrt{\mathbb{C}}}\mathbb{S}\frac{1}{\sqrt{\mathbb{L}}_x}\Pi_{N-1}. 
\end{align}
Equivalently, these conditions mean no resistive losses, $\omega_a=\omega_{\bar{a}}$, and $C_aL_{x,a}=C_{\bar{a}}L_{x,\bar{a}}$. Similarly, requiring $\mathcal{PT}$ symmetry for the anti-Hermitian potential implies $\mathcal{P}A\mathcal{P}=A$ or equivalently, $C_a=C_{\bar{a}}$, $L_a=L_{\bar{a}}$, and $L_{x,a}=L_{x,N-a}$.

Next, we outline the mapping of this $N_e$-dimensional dynamical system onto an $N$-dimensional tight-binding model. Since $H_0$ has purely imaginary entries, for a $\mathcal{PT}$-symmetric Hamiltonian, the unitary $\mathcal{U}$ in Eq.~(\ref{eq:ptne}) anticommutes with it. Therefore, the eigenvalues of $H_0$ occur in pairs $\pm\epsilon_a$, or equivalently, it has a chiral symmetry. It also follows, most clearly from Eq.~(\ref{Eq_H0G}), that $H_0$ has $(N-1)$ linearly-dependent columns and therefore $(N-1)$ eigenvalue zeros. This combination of chiral symmetry and zero modes is instrumental to mapping the $N_e=(3N-1)$ dimensional system onto an $N$-dimensional model. These arguments remain valid when a $\mathcal{PT}$-symmetric complex gauge potential $\Gamma$ is added, and therefore, the $\mathcal{PT}$-symmetric circuit Hamiltonian $H_\mathrm{circ}$ of size $N_e$ can always be mapped onto a $\mathcal{PT}$-symmetric tight-binding model with $N$ sites.  

The canonical model with an exceptional point of order $N$ is $H(\gamma)=\kappa J_x+i\gamma J_z$ where $J_x,J_z$ are $N$ dimensional representations of su(2)~\cite{Joglekar2013,Teimourpour2018,Quiroz-Juarez2019p862,Tschernig2018p1985}. However, because we have a classical system with a purely real state vector, we use its counterpart with purely imaginary entries,  $H_N(\gamma)=\kappa J_y+i\gamma J_z$. In such tight-binding lattice, the Hermitian coupling between adjacent sites is given by $J_y(a,a+1)=i\sqrt{a(N+1-a)}/2=-J_y(a+1,a)$. 
Now, to connect the $J_y$ matrix elements to our coupled electrical circuits model, we use the expression for the effective, dimensionless tunneling amplitude between two inductively coupled circuits, $J_\textrm{eff}=M^2/2\sqrt{1+M^2}$ where $M^2=L/L_{x}$~\cite{Leon-Montiel2018}. 

\begin{figure}[b!]
\includegraphics[width=1\linewidth]{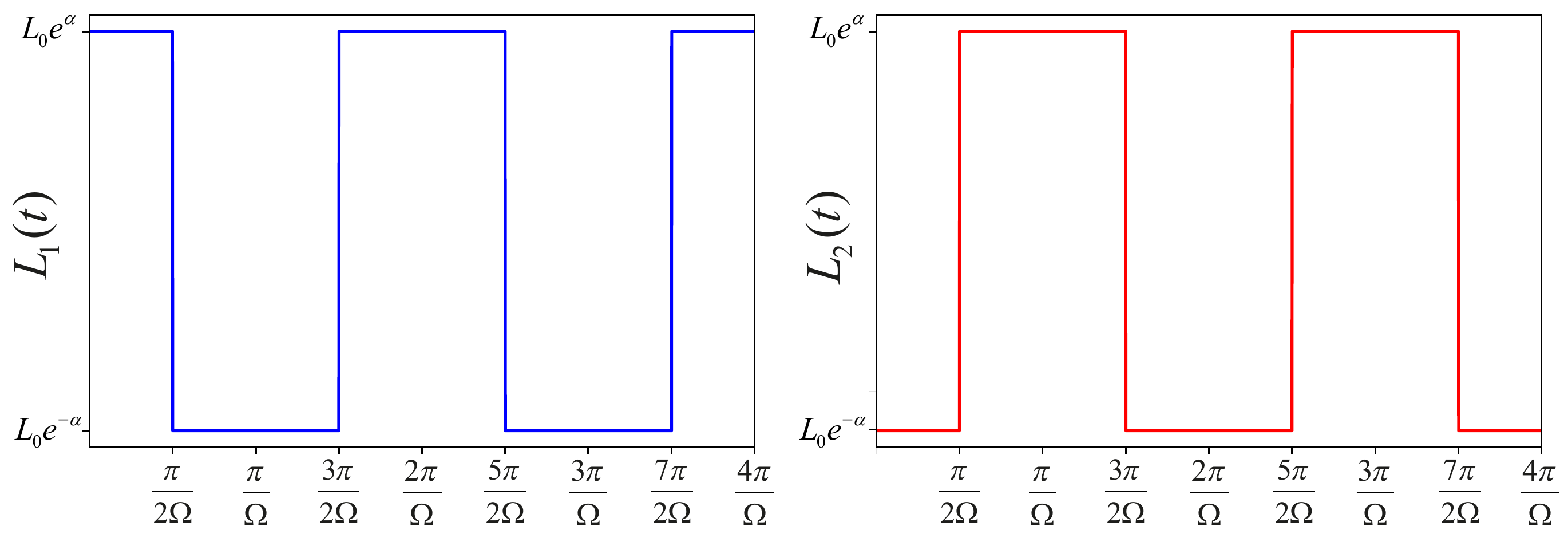}
\caption{\label{fig:Funtion_Rec} \textbf{Square-wave periodic function with modulation frequency $\Omega$}. (a) For $N=2$, the time-dependent inductances are given by $L_{1}\left( t \right)=L_{0} e^{f \left( t \right)}$. (b) shows the complementary, $L_{2}\left( t \right)=L_{0} e^{-f \left( t \right)}=L_0^2/L_1(t)$.}
\end{figure}


\begin{figure*}[t!]
\includegraphics[width=1\linewidth]{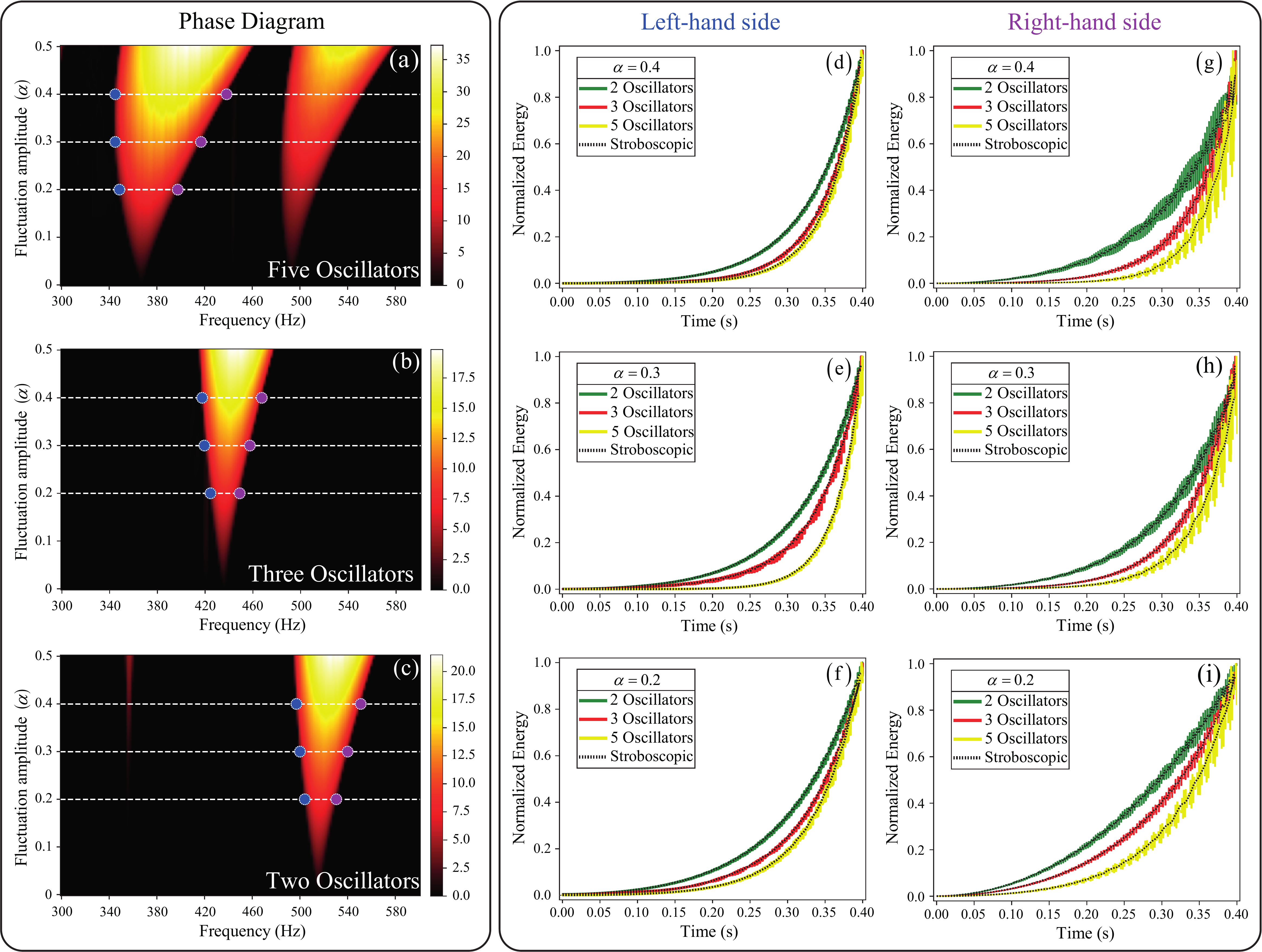}
\caption{\label{fig:High order_alpha02} \textbf{High-order EP contours in the $(\alpha,\Omega)$ plane}. (a)-(c) $\mu(\alpha,\omega)$ in the frequency window from $\Omega/(2\pi)=300$ Hz  to 600 Hz is obtained using $\tau=100$ ms. The EP contours, separating regions from light windows, are clearly seen, with EPs along the $\alpha=\{0.2,0.3,0.4\}$ lines shown as blue (left) and purple (right) filled circles. (d)-(f) normalized energy $E(t)/E_{\max}$ for blue EPs on the left-side contour shows power-law in time growth, with an exponent determined by $N$, but independent of $\alpha$. Results for two, three, and five oscillators are shown in green, red, and yellow lines, respectively; overlaid black-dashed lines show stroboscopic results. (g)-(i) corresponding results for the purple EPs on the right-side contour, shown with the same conventions, are quantitatively similar. }
\end{figure*}


Similarly, to create the gain-loss term $i\gamma(t)J_z$ the inductors $L_a(t)$ within each circuit are modulated while keeping the capacitors and coupling inductors static across the array. Since $J_z=\textrm{diag}(s,s-1,\cdots,-s)$ where $s=(N-1)/2$ is the spin associated with the $N$-dimensional representation, using the modulation
\begin{align}
    L_a(t)\equiv L_0e^{f_a(t)}=L_0 e^{(N+1-2a)\int_0^t\gamma(t')dt'},
    \label{eq:la}
\end{align}
leads to $\Gamma=i\gamma(t)\textrm{diag}(0_N,J_z,0_{N-1})$. This temporal variation means the inductors in mirror-symmetric positions are varied in an inverse manner,  $L_a(t)L_{\bar{a}}(t)=L_0^2=\textrm{const}$. Therefore, when inductance increase at site $a$, indicating gain, is balanced by inductance decrease at its mirror symmetric site $\bar{a}$, indicating loss, and $\mathcal{PT}$ symmetry can be created without real amplifying or Joule-heating elements. The exponent function $f(t)=\int\gamma(t')dt'$ in Eq.~(\ref{eq:la}) allows us to create arbitrary, balanced gain-loss profiles. 

The choice of $f(t)$ is informed by the ability to dynamically modulate the synthetic inductances in real time by using electronic circuits \cite{Quiroz-Juarez2021parXiv}. Creating a static gain-loss term requires inductances that either grow or decay exponentially~\cite{Quiroz-Juarez2021parXiv}. However, Floquet EP contours of the same order also emerge by periodic variations~\cite{Joglekar20114,Leon-Montiel2018,Quiroz-Juarez2021parXiv}. For simplicity, we consider the square-wave function
\begin{eqnarray}  \label{Square_funtion}
	f\left(t\right)=\left\{\begin{array}{cc}
 \alpha & 0\leq t\leq T/4,\\
 -\alpha & T/4< t< 3T/4,\\
 \alpha & 3T/4 < t\leq T,
 \end{array}\right.
\end{eqnarray}
where $T$ is the period and $\Omega=2\pi/T$ defines the modulation frequency. For example, when $N=2$, the two inductances satisfy $e^{- \alpha}\leq L_{1,2}\left( t \right)/L_0\leq e^{\alpha}$ (Fig.~\ref{fig:Funtion_Rec}). 
Since the inductance in each RLC box varies with time, the Hamiltonian $H_0$ also acquires time-dependence through the matrix $\mathbb{W}$ of frequencies $\omega_a(t)=1/\sqrt{L_a(t)C_a}$.

The $\mathcal{PT}$ phase diagram of this system can be obtained via two methods. The first uses the non-unitary time evolution operator $G_F(T)$ to calculate the equivalent non-Hermitian, $\mathcal{PT}$-symmetric Floquet Hamiltonian $G_F(T))\equiv\exp[-iH_F(\alpha,\Omega)T]$~\cite{Joglekar20114}. The second, experimentally friendly approach is to obtain the time-dependent circuit energy $E(t)$, Eq.~(\ref{eq:energy}), and compare its growth over sufficiently long time-intervals $\tau$ and $2\tau$. To quantify it, we define a dimensionless ratio~\cite{Leon-Montiel2018,Quiroz-Juarez2021parXiv} 
\begin{eqnarray}  \label{mu}
	\mu =\log \left\{ \frac{\max \left[ E\left( 0\le t\le 2\tau  \right) \right]}{\max \left[ E\left( 0\le t\le \tau  \right) \right]} \right\}.
\end{eqnarray}
In the $\mathcal{PT}$-symmetric phase with time-periodic dynamics, $\max E(t)$ will be the same over the two intervals, and therefore $\mu=0$ denotes the $\mathcal{PT}$-symmetric phase. On the other hand, in $\mathcal{PT}$-broken phase with exponentially amplifying modes, Eq.~(\ref{mu}) provides a linear-in-$\tau$ metric that indicates the average amplification $\mu>0$. The $\mathcal{PT}$ transition is accompanied by a vanishing energy gap and divergent period on the $\mathcal{PT}$-symmetric side of the boundary. Therefore, at any finite $\tau$, this approach leads to some smearing of the EP contours. In the following section, we present the results of such an analysis. 


\section{Results}
\label{sec:floquet}

For numerical analysis, we use experimentally accessible and viable circuit parameters~\cite{Quiroz-Juarez2021parXiv}: resistance $R_0=1$ k$\Omega$, inductance $L_{0}=0.01$ H and capacitance $C_{0}=100$ $\mu$F. Thus, the isolated oscillator frequency is $\omega_{0}/(2\pi)=159.15$ Hz, and the isolated oscillator RC-decay rate, $1/R_0C_0=10$ Hz, is much smaller than the natural frequency, thereby justifying the approximation $\mathbb{G}_{RC}\approx 0$. The coupling inductances are set to $L_{x1}=0.5L_{0}$ for $N=2$ oscillators, $L_{x1}=L_{x2}=0.5L_{0}$ for $N=3$ oscillators, and $L_{x1}=L_{x4}=0.67L_{0}$, $L_{x2}=L_{x3}=0.5L_{0}$ for $N=5$ as necessitated by the non-uniform matrix elements of the $J_y$ array. We use $\alpha\leq 0.4$, meaning the inductances span $0.67L_0\leq L_{a}(t)\leq 1.5L_0$, a range that can be dynamically achieved in the synthetic circuits~\cite{Leon-Montiel2018,Quiroz-Juarez2021parXiv}. 

The left-hand panel in Fig.~\ref{fig:High order_alpha02} shows the numerically computed $\mu(\alpha,\Omega)$ for $N=5$ (a), $N=3$ (b), and $N=2$ (c) over a modulation-frequency window $\Omega/(2\pi)$ from 300 Hz to 600 Hz. The dark regions with $\mu=0$ denote the $\mathcal{PT}$-symmetric phase, where the circuit energy $E(t)$ undergoes bounded oscillations. They are punctuated by bright, triangular, $\mathcal{PT}$-symmetry broken regions ($\mu>0$) that occur down to vanishingly small non-Hermiticity $\alpha\ll 1$ at specific frequencies~\cite{Joglekar20114,Leon-Montiel2018,Quiroz-Juarez2021parXiv}. These regions are separated by EP contours with order $N$. 

We analyze the dynamics at the EPs by monitoring how fast the circuit energy $E(t)$ increases when the system is parked at the EPs (shown by blue, left and purple, right circles with white boundaries) along $\alpha=\{0.2,0.3,0.4\}$ lines. The center panel in Fig.~\ref{fig:High order_alpha02} shows normalized energies $E\left(t\right)/E_{\max}$ (solid lines) and their respective stroboscopic results (black-dashed lines) at $\alpha=0.4$ (d), $\alpha=0.3$ (e), and $\alpha=0.2$ (f) when the system is parked on the blue, left EPs. Figure~\ref{fig:High order_alpha02} (g)-(i) show corresponding results when the system is parked on the purple, left EPs. 
In each case, it can be seen that $E(t)/E_{\max}$ follows a power-law dependence on $t$ with an exponent that increases with the order of the EP.
This result is independent of the degree of non-Hermiticity $\alpha$ or the location--left or right---of the EP contour. The large fluctuations in the non-stroboscopic data for $E(t)$ also hint at the asymmetric dependence of $E(t)$ on the EP location at a fixed $\alpha$~\cite{Quiroz-Juarez2021parXiv}.

\begin{figure}
\includegraphics[width=1\linewidth]{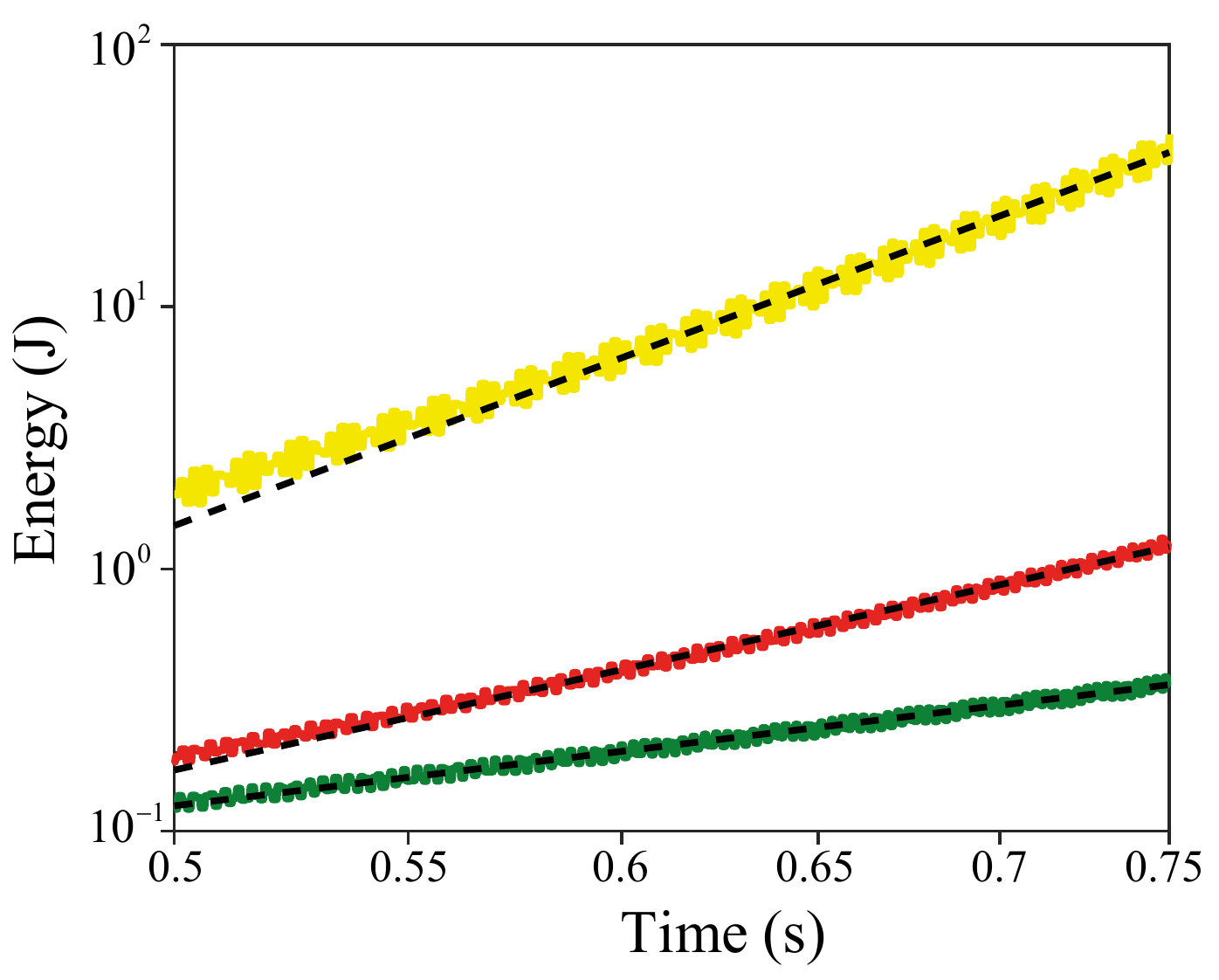}
\caption{\label{fig:Fitting_Energy} \textbf{Energy increase at the EPs}. Unnormalized circuit energy $E(t)$ on a log-log scale at long times shows linear behavior with a slope that depends on the order $N$ of the EP. The results are for $\alpha=0.2$, purple (right-side) EP. Similar results are obtained, with appropriate long-time windows, for left- and right-side EP contours for all $\alpha$. For the representative data shown, the power-law exponents are $2.62$, $4.83$, and $8.08$, for $N=2$, $N=3$, and $N=5$, respectively.}
\end{figure}

Since the system with $N$ has an $N$th order EP, we expect the unnormalized energy to grow as $E(t)\propto t^{2(N-1)}$ at long times. To confirm this, Figure~\ref{fig:Fitting_Energy} shows the system's energy vs time on a log-log scale for $\alpha=0.2$. The slopes obtained from straight-line fits to the log-log data are in agreement with the prediction that the power-law exponent for an EP of order $N$ is given by $2(N-1)$~\cite{Quiroz-Juarez2019p862,Quiroz-Juarez2021parXiv}.

\section{Conclusion}
\label{sec:disc}
From their start in quantum theory and mathematical physics, non-Hermitian, $\mathcal{PT}$-symmetric models are now studied across the board in fields as widely varied as minimal quantum systems or a single LC oscillator. This veritable cornucopia of experimental realizations has also invited detailed comparison of seemingly different models. Here, we have analyzed one such model, an array of $N$ inductively coupled RLC circuits with dynamic parameters to show that energy dynamics in it is generated by a $(3N-1)$-dimensional non-Hermitian Hamiltonian, and through general formalism, spelled out the constraints that make such Hamiltonian have chiral and $\mathcal{PT}$ symmetry. We have then shown that this model, with $2N$ dynamical variables, can be mapped onto an $N$-dimensional tight-binding lattice that can support an EP of order $N$. 

By implementing the gain and loss through a periodic variation of the inductances in the RLC units, we have numerically mapped out the Floquet $\mathcal{PT}$-phase diagram for two, three, and five oscillator chains, all of which show EP contours at vanishingly small non-Hermiticities. By tracking the circuit's energy, we are able extract the order of EP by looking at the power-law-in-time exponent for the $E(t)$ increase. Our results will be useful for realizing robust, arbitrary-order EPs by means of complex gauge fields in dynamically modulated synthetical oscillator networks. 


This work was supported by DGAPA-UNAM under the project UNAM-PAPIIT IN101623, and by CONACYT under the project No. A1-S-8317. J.D.H.-M. thankfully acknowledges financial support by CONACYT. Y.N.J. is supported by ONR Grant No. N00014-21-1-2630.


\appendix

\section{Explicit Hamiltonian expressions for two, three, and five oscillator circuits.} 

\subsection{Two coupled RLC oscillators}
For two coupled RLC oscillators, $\ket{ \phi \left(t\right)}=\left( V_1,V_2,I_1,I_2,I_{x1} \right)^{T}$ and the $5\times 5$ matrix $H\left(t\right)$ with purely imaginary entries is given by 
\begin{eqnarray}
	H \left(t\right)=i\left[\begin{array}{ccccc}
		-\frac{1}{C_{1} R_{1}} & 0 & -\frac{1}{C_{1}} & 0 & -\frac{1}{C_{1}} \\ 0 & -\frac{1}{C_{2} R_{2}} & 0 & -\frac{1}{C_{2}} & \frac{1}{C_{2}} \\ \frac{1}{L_{1} \left(t \right)} & 0 & 0 & 0 & 0\\ 0 & \frac{1}{L_{2} \left(t \right)} & 0 & 0 & 0 \\ \frac{1}{L_{12}} & -\frac{1}{L_{12}} & 0 & 0 & 0
	\end{array}\right].  \nonumber \\
\end{eqnarray}
To generate balanced gain-loss through the temporal modulation of the inductors, we use $L_{1}\left(t\right)=L_{0} e^{ f\left(t\right)}$ and $L_{2}\left(t\right)=L_{0} e^{- f\left(t\right)}$, and set $C_{1,2}=C_{0}$, $L_{12}=0.5L_{0}$ and $R_{1,2}=R_0$. With $\gamma(t)=df/dt$, the circuit Hamiltonian can be written as
\begin{eqnarray}  
	H_{\textrm{cir}}\left(t\right) =i\left[\begin{array}{ccccc}
		-\frac{1}{C_0R_0} & 0 & -\frac{\omega_{0}}{e^{f\left(t\right)/2}}  & 0 & -\frac{\omega_{0}}{\sqrt{0.5} }  \\ 
        0 & -\frac{1}{C_0R_0} & 0 & -\frac{\omega_{0}}{e^{-f\left(t\right)/2}}   & \frac{\omega_{0}}{\sqrt{0.5}}   \\ 
        \frac{\omega_{0}}{e^{f\left(t\right)/2}} & 0 & \frac{\gamma\left(t\right)}{2} & 0 & 0\\ 
        0 & \frac{\omega_{0}}{e^{- f\left(t\right)/2}}  & 0 & -\frac{\gamma\left(t\right)}{2} & 0 \\ 
        \frac{\omega_{0}}{\sqrt{0.5} }  & -\frac{\omega_{0} }{\sqrt{0.5}} & 0 & 0 & 0
		\end{array}\right]. \nonumber  \\
\end{eqnarray}


\subsection{Three coupled RLC oscillators}
For three coupled RLC oscillators, $\ket{ \phi \left(t\right) }=\left(  V_1,V_2,V_3,I_1,I_2,I_3,I_{x1},I_{x2} \right)^{T}$, and the $8\times 8$ matrix $H(t)$ with purely imaginary entries is given by 
\begin{widetext}
\begin{eqnarray}
	H\left(t\right)= i\left[\begin{array}{cccccccc}
		-\frac{1}{C_{1} R_{1}} & 0 & 0 & -\frac{1}{C_{1}} & 0 & 0 & -\frac{1}{C_{1}} & 0  \\ 0 & -\frac{1}{C_{2} R_{2}} & 0 & 0 & -\frac{1}{C_{2}} & 0 & \frac{1}{C_{2}} & -\frac{1}{C_{2}} \\ 0 & 0 & -\frac{1}{C_{3} R_{3}} & 0 & 0 & -\frac{1}{C_{3}} & 0 & \frac{1}{C_{3}}  \\ \frac{1}{L_{1} \left(t \right)} & 0 & 0 & 0 & 0 & 0 & 0 & 0 \\ 0 & \frac{1}{L_{2}} & 0 & 0 & 0 & 0 & 0 & 0   \\ 0 & 0 & \frac{1}{L_{3} \left(t \right)} & 0 & 0 & 0 & 0 & 0   \\ \frac{1}{L_{x1}} & -\frac{1}{L_{x1}} & 0 & 0 & 0 & 0 & 0 & 0  \\ 0 & \frac{1}{L_{x2}} & -\frac{1}{L_{x2}} & 0 & 0 & 0 & 0 & 0 
	\end{array}\right].
\end{eqnarray}
\end{widetext}
To maintain $\mathcal{PT}$-symmetry and generate balanced gain and loss, we use $L_{x1}=L_{x2}=0.5L_{0}$, $L_{1}\left(t\right)=L_{0} e^{f\left(t\right)}$, $L_{2}=L_{0}$ and $L_{3}\left(t\right)=L_{0} e^{-f\left(t\right)}$. The resulting circuit Hamiltonian becomes

\begin{widetext}
\begin{eqnarray}   
\label{Heff_three}
H_\textrm{cir}\left(t\right)=i\left[
\begin{array}{cccccccc}
-\frac{1}{C_0R_0} & 0 & 0 & -\frac{\omega_0}{e^{f \left( t \right)/2}}  & 0 & 0 & -\frac{\omega_0}{\sqrt{0.5}}  & 0   \\
0 & -\frac{1}{C_0R_0} & 0 & 0 & -\omega_0 & 0 & \frac{\omega_0}{\sqrt{0.5} } & -\frac{\omega_0}{\sqrt{0.5}}   \\
0 & 0 & -\frac{1}{C_0R_0} & 0 & 0 & -\frac{\omega_0}{e^{-f\left( t \right)/2 }}  & 0 & \frac{\omega_0}{ \sqrt{0.5} }  \\
\frac{\omega_0}{e^{f \left( t \right)/2 }}  & 0 & 0 & \gamma\left(t\right) & 0 & 0 & 0 & 0 \\
0 & \omega_0 & 0 & 0 & 0 & 0 & 0 & 0  \\
0 & 0 & \frac{\omega_0 }{e^{- f \left( t \right)/2 }} & 0 & 0 & -\gamma\left(t\right) & 0 & 0  \\
\frac{\omega_0 }{\sqrt{0.5}}  & -\frac{\omega_0 }{\sqrt{0.5}} & 0 & 0 & 0 & 0 & 0 & 0  \\
0 & \frac{\omega_0 }{\sqrt{0.5}} & -\frac{\omega_0}{\sqrt{0.5}}  & 0 & 0 & 0 & 0 & 0  
\end{array}
\right].
\end{eqnarray}
\end{widetext}


    
\subsection{Five coupled $RLC$ oscillators}
For $N-5$, the 14-dimensional state vector is $\ket{\phi(t)}=\left(V_1,V_2,V_{3},V_{4},V_{5},I_1,I_2,I_{3},I_{4},I_{5},I_{x1},I_{x2},I_{x3},I_{x4} \right)^{T}$ and the 14-dimensional, purely imaginary matrix $H \left( t \right)$ becomes

\begin{widetext}
{\footnotesize
\begin{eqnarray}
		H\left(t\right)=i\left[
		\begin{array}{cccccccccccccc}
			-\frac{1}{C_1 R_1} & 0 & 0 & 0 & 0 & -\frac{1}{C_1} & 0 & 0 & 0 & 0 & -\frac{1}{C_1} & 0 & 0 & 0 \\
			0 & -\frac{1}{C_2 R_2} & 0 & 0 & 0 & 0 & -\frac{1}{C_2} & 0 & 0 & 0 & \frac{1}{C_2} & -\frac{1}{C_2} & 0 & 0 \\
			0 & 0 & -\frac{1}{C_3 R_3} & 0 & 0 & 0 & 0 & -\frac{1}{C_3} & 0 & 0 & 0 & \frac{1}{C_3} & -\frac{1}{C_3} & 0 \\
			0 & 0 & 0 & -\frac{1}{C_4 R_4} & 0 & 0 & 0 & 0 & -\frac{1}{C_4} & 0 & 0 & 0 & \frac{1}{C_4} & -\frac{1}{C_4} \\
			0 & 0 & 0 & 0 & -\frac{1}{C_5 R_5} & 0 & 0 & 0 & 0 & -\frac{1}{C_5} & 0 & 0 & 0 & \frac{1}{C_5} \\
			\frac{1}{L_1 \left(t\right)} & 0 & 0 & 0 & 0 & 0 & 0 & 0 & 0 & 0 & 0 & 0 & 0 & 0 \\
			0 & \frac{1}{L_2 \left(t\right)} & 0 & 0 & 0 & 0 & 0 & 0 & 0 & 0 & 0 & 0 & 0 & 0 \\
			0 & 0 & \frac{1}{L_3} & 0 & 0 & 0 & 0 & 0 & 0 & 0 & 0 & 0 & 0 & 0 \\
			0 & 0 & 0 & \frac{1}{L_4 \left(t\right)} & 0 & 0 & 0 & 0 & 0 & 0 & 0 & 0 & 0 & 0 \\
			0 & 0 & 0 & 0 & \frac{1}{L_5 \left(t\right)} & 0 & 0 & 0 & 0 & 0 & 0 & 0 & 0 & 0 \\
			\frac{1}{L_{x1}} & -\frac{1}{L_{x1}} & 0 & 0 & 0 & 0 & 0 & 0 & 0 & 0 & 0 & 0 & 0 & 0 \\
			0 & \frac{1}{L_{x2}} & -\frac{1}{L_{x2}} & 0 & 0 & 0 & 0 & 0 & 0 & 0 & 0 & 0 & 0 & 0 \\
			0 & 0 & \frac{1}{L_{x3}} & -\frac{1}{L_{x3}} & 0 & 0 & 0 & 0 & 0 & 0 & 0 & 0 & 0 & 0 \\
			0 & 0 & 0 & \frac{1}{L_{x4}} & -\frac{1}{L_{x4}} & 0 & 0 & 0 & 0 & 0 & 0 & 0 & 0 & 0 \\
		\end{array}
		\right].
\end{eqnarray}}
To create $\mathcal{PT}$-symmetric circuit with balanced gain and loss, we use $L_{x1}=0.67L_{0}=L_{x4}$, $L_{x2}=0.5L_{0}=L_{x3}$. The time-dependent inductors are vary as $L_{1,5}\left(t\right)=L_{0} e^{\pm 2 f\left(t\right)}$, $L_{2,4}\left(t\right)=L_{0} e^{\pm  f\left(t\right)}$, and $L_{3}=L_{0}$. This leads to a $14\times 14$ circuit Hamiltonian with purely imaginary entries 
{\footnotesize{
		\begin{eqnarray} 
			H_{\textrm{cir}}\left(t\right)=i\left[
			\begin{array}{cccccccccccccc}
				-\frac{1}{C_0R_0} & 0 & 0 & 0 & 0 & -\frac{\omega _0}{e^{f \left(t\right)}} & 0 & 0 & 0 & 0 &  -\frac{\omega_{0}}{\sqrt{0.67}} & 0 & 0 & 0 \\
				0 & -\frac{1}{C_0R_0} & 0 & 0 & 0 & 0 & -\frac{\omega _0 }{e^{ f \left(t\right)/2}} & 0 & 0 & 0 & \frac{\omega_{0}}{\sqrt{0.67}} & -\frac{\omega_{0}}{\sqrt{0.5}} & 0 & 0 \\
				0 & 0 & -\frac{1}{C_0R_0} & 0 & 0 & 0 & 0 & -\omega _0 & 0 & 0 & 0 &  \frac{\omega_{0}}{\sqrt{0.5}} &  -\frac{\omega_{0}}{\sqrt{0.5}} & 0 \\
				0 & 0 & 0 & -\frac{1}{C_0R_0} & 0 & 0 & 0 & 0 & -\frac{\omega _0}{e^{-f \left(t\right)/2}} & 0 & 0 & 0 &  \frac{\omega_{0}}{\sqrt{0.5}} &  -\frac{\omega_{0}}{\sqrt{0.67}} \\
				0 & 0 & 0 & 0 & -\frac{1}{C_0R_0} & 0 & 0 & 0 & 0 & -\frac{\omega_0 }{e^{-f \left(t\right)}}  & 0 & 0 & 0 &  \frac{\omega_{0}}{\sqrt{0.67}} \\
				\frac{\omega_0}{e^{f \left(t\right)} }& 0 & 0 & 0 & 0 & 2\gamma (t) & 0 & 0 & 0 & 0 & 0 & 0 & 0 & 0 \\
				0 & \frac{\omega_0}{e^{f \left(t\right)/2}}  & 0 & 0 & 0 & 0 & \gamma (t) & 0 & 0 & 0 & 0 & 0 & 0 & 0 \\
				0 & 0 & \omega_0 & 0 & 0 & 0 & 0 & 0 & 0 & 0 & 0 & 0 & 0 & 0 \\
				0 & 0 & 0 & \frac{\omega_0}{e^{-f \left(t\right)/2}}  & 0 & 0 & 0 & 0 & -\gamma (t) & 0 & 0 & 0 & 0 & 0 \\
				0 & 0 & 0 & 0 & \frac{\omega_0}{e^{-f \left(t\right)}}  & 0 & 0 & 0 & 0 & -2\gamma (t) & 0 & 0 & 0 & 0 \\
				 \frac{\omega_{0}}{\sqrt{0.67}} & -\frac{\omega_{0}}{\sqrt{0.67}} & 0 & 0 & 0 & 0 & 0 & 0 & 0 & 0 & 0 & 0 & 0 & 0 \\
				0 &  \frac{\omega_{0}}{\sqrt{0.5}} & -\frac{\omega_{0}}{\sqrt{0.5}} & 0 & 0 & 0 & 0 & 0 & 0 & 0 & 0 & 0 & 0 & 0 \\
				0 & 0 & \frac{\omega_{0}}{\sqrt{0.5}} & -\frac{\omega_{0}}{\sqrt{0.5}} & 0 & 0 & 0 & 0 & 0 & 0 & 0 & 0 & 0 & 0 \\
				0 & 0 & 0 & \frac{\omega_{0}}{\sqrt{0.67}} & -\frac{\omega_{0}}{\sqrt{0.67}} & 0 & 0 & 0 & 0 & 0 & 0 & 0 & 0 & 0 \\
			\end{array}
			\right]. \nonumber  \\ 
\end{eqnarray}}}
\end{widetext}



\bibliography{biblio}

\end{document}